\documentclass[aoas,MSNbibl,nameyear,seceqn,dvips]{arximspdf}
\usepackage{graphicx}
\doi{10.1214/13-AOAS681} 
\volume{7}
\issue{4}
\pubyear{2013}
\firstpage{2007}
\lastpage{2033}

\makeatletter
\newtheorem{theorem}{Theorem}
\newtheorem{lemma}{Lemma}
\newtheorem{derivation}[theorem]{Derivation} 
\newproclaim{definition}{Definition}
\newproclaim{rems}{Remarks}
\makeatother

\begin{document}
\begin{frontmatter}

\title{The shuffle estimator for explainable variance in~fMRI experiments\thanksref{T1}}
\runtitle{Shuffle estimator in fMRI}
\pdftitle{The shuffle estimator for explainable variance in fMRI experiments}

\begin{aug}
\author[A]{\fnms{Yuval} \snm{Benjamini}\corref{}\ead[label=e1]{yuvalben@stanford.edu}}
\and
\author[B]{\fnms{Bin} \snm{Yu}\ead[label=e2]{binyu@stat.berkeley.edu}}
\runauthor{Y. Benjamini and B. Yu}
\affiliation{Stanford University and University of California, Berkeley}
\address[A]{Department of Statistics\\
Stanford University\\
Stanford, California 94305\\
USA\\
\printead{e1}}
\address[B]{Department of Statistics\\
University of California, Berkeley\\
367 Evans Hall\\
Berkeley, California 94720\\
USA\\
\printead{e2}}
\end{aug}
\thankstext{T1}{Supported in part by NSF Grants DMS-11-07000,
SES-0835531 (CDI), ARO Grant W911NF-11-1-0114 and the Center for
Science of
Information (CSoI) and NSF Science and Technology Center, under
Grant agreement CCF-0939370.}

\received{\smonth{5} \syear{2012}}
\revised{\smonth{6} \syear{2013}}

%
\begin{abstract}
In computational neuroscience, it is important to estimate well the
proportion of signal variance in the total variance of neural activity
measurements. This explainable variance measure helps neuroscientists
assess the adequacy of predictive models that describe how images are
encoded in the brain. Complicating the estimation problem are strong
noise correlations, which may confound the neural responses
corresponding to the stimuli. If not properly taken into account, the
correlations could inflate the explainable variance
estimates and suggest false possible prediction accuracies.

We propose a novel method to estimate the explainable variance in
functional MRI (fMRI) brain activity measurements when there are strong
correlations in the noise. Our shuffle estimator is nonparametric,
unbiased, and built upon the random effect model reflecting the
randomization in the fMRI data collection process. Leveraging
symmetries in the measurements, our estimator is obtained by
appropriately permuting the measurement vector in such a way that the
noise covariance structure is intact but the explainable variance is
changed after the permutation. This difference is then used to estimate
the explainable variance. We validate the properties of the proposed
method in simulation experiments. For the image-fMRI data, we show that
the shuffle estimates can explain the variation in prediction accuracy
for voxels within the primary visual cortex (V1) better than
alternative parametric methods.
\end{abstract}

%
\begin{keyword}
\kwd{Analysis-of-variance}
\kwd{encoding models}
\kwd{explainable variance}
\kwd{fMRI}
\kwd{permutations}
\kwd{prediction}
\kwd{random-effects}
\kwd{shuffle estimator}
\kwd{vision}
\end{keyword}
\pdfkeywords{Analysis-of-variance, encoding models, explainable variance, fMRI, permutations, prediction, random-effects, shuffle estimator, vision}

\end{frontmatter}

\setcounter{footnote}{1}
\section{Introduction}
Neuroscientists study how human perception of the outside world is
physically encoded in the brain. Although the brain's processing unit,
the neuron, performs simple manipulations of its inputs, hierarchies of
interconnected neuron groups achieve complex perception tasks. By
measuring neural activities at different locations in the hierarchy,
scientists effectively sample different stages in the cognitive
process.

Functional MRI (fMRI) is an indirect imaging technique, which allows
researchers to sample a correlate of neural activities over a dense
grid covering the brain. FMRI measures changes in the magnetic field
caused by flow of oxygenated blood; these blood oxygen-level dependent
(BOLD) signals are indicative of neuronal activities. Because it is
noninvasive, fMRI can record neural activity from a human subject's
brain while the subject performs cognitive tasks that range from basic
perception of images or sound to higher-level cognitive and motor
actions. The vast data collected by these experiments allow
neuroscientists to develop quantitative models, \emph{encoding models}
[\citet{Dayan}], that relate the cognitive tasks with the activity
patterns these tasks evoke in the brain. Encoding models are usually
fit separately to each point of the spatial activity grid, a
\emph{voxel}, recorded by fMRI. Each fitted encoding model extracts
features of the perceptual input and summarizes them into a value
reflecting the evoked activity at the voxel.

Encoding models are important because they can be quantitatively
evaluated based on how well they can predict on new data. Prediction
accuracy of different models is thus a yardstick to contrast competing
models regarding the function of the neurons spanned by the voxel
[\citet{EarlyVisual}]. Furthermore, the relation between the
spatial organization of neurons along the cortex and the function of
these neurons can be recovered by feeding the model with artificial
stimuli. Finally, predictions for multiple voxels taken together create
a predicted fingerprint of the input; these fingerprints have been
successfully used for extracting information from the brain [so-called
``mind-reading'', \citet{Shinji2011}] and building brain machine
interfaces [\citet{ShohamBMI}]. The search for simpler but more
predictive encoding models is ongoing, as researchers try to encode
more complex stimuli and predict higher levels of cognitive processing.

Because brain responses are not deterministic, encoding models cannot
be perfect. A substantial portion of the fMRI measurements is noise
that does not reflect the input. The noise may be caused by background
brain activity, by noncognitive factors related to blood circulation or
by the measurement apparatus. Regardless of the source, the noise
cannot be predicted by encoding models that are deterministic functions
of the inputs [\citet{Assessing}]. To reduce the effect of noise,
the same input can be displayed multiple times within the input
sequence and all responses to the same input averaged, in an
experimental design called event-related fMRI
[\citet{EventRelated}]. See \citet{PasupathyA}, \citet{HaefnerC}, for
examples, and \citet{Huettel} for a review. Typically, even after
averaging, the noise level is high enough to be a considerable source
of prediction error. Hence, it is standard practice to measure and
report an indicator of the signal strength together with prediction
success. We focus on \emph{explainable variance}, the proportion of
signal variance in the total variance of the
measurements,\footnote{Depending on context, this proportion is also
known as the intraclass correlation, effect-size and~$R^2$.} which we here formally define in equation (\ref{EVdef}). The
comparison of explainable variance with prediction success
[\citet{Assessing}, \citet{SahaniAndLinden}] informs how much room is left
on the collected data for improving prediction through better models.
Explainable variance is also an important quality control metric before
fitting encoding models, and can help choose regularization parameters
for model training.

In this paper we develop a new method to estimate the explainable
variance in fMRI responses and use it to reanalyze data from an
experiment conducted by the Gallant lab at UC Berkeley
[\citet{Kayetall}, \citet{Naselarisetall}]. Their work examines the
representation of visual inputs in the human brain using fMRI by
ambitiously modeling a rich class of images from natural scenes rather
than artificial stimulus. An encoding model was fit to each of more
than 10,000 voxels within the visual cortex. The prediction accuracy of
their fitted models on a separate validation image set was surprisingly
high given the richness of the input class, inspiring many studies of
rich stimuli class encoding [\citet{ReconSpeech}, \citet{GeneratingText}].
Still, accuracy for the voxels varied widely (see Figure~\ref{DataEA}),
and more than a third of the voxels had prediction accuracy not
significantly better than random guessing. Researchers would like to
know whether accuracy rates reflect (a) overlooked features which might
have improved the modeling, or instead reflect (b) the noise that
cannot be predicted regardless of the model used. As we show in this
paper, valid measures of explainable variance can shed light on this
question.

\subsection*{Measuring explainable variance on correlated noise}
We face the statistical problem of estimating the explainable variance,
assuming the measurement vector is composed of a random mean-effects
signal evoked by the images with additive auto-correlated noise
[\citet{ScheffeBook}]. In fMRI data, many of the sources of noise
would likely affect more than one measurement. Furthermore, low
frequency correlation in the noise has been shown to be persistent in
fMRI data [\citet{LFF}]. Ignoring the correlation would greatly
bias the signal variance estimation (see Figure~\ref{OptvsObs} below)
and would cause us to overestimate the explainable variance. This
overestimation of signal variance may be a contributing factor to
replicability concerns raised in neuroscience [\citet{Voodoo}].

Classical analysis-of-variance methods account for correlated noise by
(a)~estimating the full noise covariance, and (b) deriving the
variances of the signal and the averaged noise based on that
covariance. The two steps can be performed separately by methods
of\vadjust{\goodbreak}
moments [\citet{ScheffeBook}] or simultaneously using restricted
maximum likelihood [\citet{REML}]. In both cases, some parametric
model for the correlation is needed for the methods to be feasible, for
example, a fast decay is assumed in \citet{TempAuto}. These
approaches, however, are sensitive to misspecifications of the
correlation parameters. In fMRI signals, the correlation of the noise
might vary with the specifics of the preprocessing method in a way,
that is, not easy to parametrize. As we show in Section~\ref{secData},
if the autocorrelation model is too simplistic, it might not capture
the correlation well and then overestimate the signal; on the other
hand, if it is too flexible, the noise might be overestimated and,
furthermore, the numeric optimizations involved in estimating the
correlation might sometimes fail to converge.

An alternative way [\citet{SahaniAndLinden}, \citet{HsuTheunissen}] to
handle the noise correlation when estimating variances is to restrict
the analysis to measurements that, based on the data collection, should
be independent. Many neuroscience experiments are divided into several
sessions, or \emph{blocks}, to better reflect the inherent variability
and to allow the subject rest. Fewer have a \emph{block design}, where
the same stimulus sequence is repeated for multiple blocks. Under block
design the signal level can be estimated by comparing repeated measures
across different blocks: regardless of the within-block-correlation,
the noise should decay as $1/b$ when averaged over $b$ blocks with the
same stimulus sequence. Block designs, however, are quite limiting for
fMRI experiments, because the long reaction time of fMRI limits the
number of stimuli that can be displayed within an experimental block
[\citet{Huettel}]. The methods above also do not use repeats
within a block to improve their estimates. These problems call for a
method that can make use of patterns in the data collection to estimate
the signal and noise variances under less restrictive designs.

We introduce novel variance estimators for the signal and noise levels,
which we call \emph{shuffle estimators}. Shuffle estimators resemble
bias correction methods: we think of
the noise component as a bias and try to remove it by resampling. %
The key idea is to artificially create a second data vector that will
have similar noise patterns as our original data. We do this by
permuting, or \emph{shuffling}, the original data along symmetries that
are based on the data collection, such as temporal stationarity or
independence across blocks. As we prove in Section~\ref{secShuffle},
the variance due to signal will be reduced in the shuffled data when
some repeated measures for the same image are shuffled into different
categories. An unbiased estimator of the signal level can be derived
based on this reduction in variance. The method does not require
parametrization of the noise correlation, and is flexible to
incorporate different structures in the data collection.

We validate our method on both simulated and fMRI data. For the fMRI
experiment, we estimate upper bounds for prediction accuracy based on
the explainable variance of each voxel in the primary visual
cortex~(V1). The upper bounds we estimate (in Section~\ref{secData})
are highly correlated ($r>0.9$) to the accuracy of the prediction
models used by the neuroscientists. We therefore postulate that
explainable variance, as estimated by the shuffle estimators, can
``predict'' optimal accuracy even for areas that do not have a good
encoding model. Alternative estimates for explainable variance showed
substantially less agreement with the prediction results of the voxels.

In Section~\ref{secPreliminaries} we describe the fMRI experiment in
greater detail and motivate the random effects model underlying our
analysis. In Section~\ref{secShuffle} we introduce the shuffle
estimators method for estimating the signal and noise levels and prove
the estimators are unbiased. In Section~\ref{secExplainable} we focus
on the relation between explainable variance and prediction for random
effects model with correlated noise. The simulations in
Section~\ref{secSimulation} verify unbiasedness of the signal estimates
for various noise regimes and show that the estimates are comparable to
parametric methods with the correct noise model. In
Section~\ref{secData} we estimate the explainable variance for multiple
voxels from the fMRI experiment and show the shuffle estimates
outperform alternative estimates in explaining variation in prediction
accuracies of the voxels. Section~\ref{secDiscussion} concludes this
paper with a discussion of our method. The proofs and the conditions of
consistency for the estimator are available in the supplementary materials [\citet{SupplementShuffle}].

\section{Preliminaries}\label{secPreliminaries}
\subsection{An fMRI experiment}
In this experiment, carried out by the Gallant lab at UC Berkeley
[\citet{Kayetall}], a human subject viewed natural images while
scanned by fMRI.\footnote{We use data from subject S1 in Kay et al. (\citeyear{Kayetall}).}
The two primary goals of the experiment were (a) to find encoding
models that have high predictive accuracy across many voxels in the
early visual areas; and (b) to use such models to identify the input
image, from a set of candidate images, based on the evoked brain
patterns. The experiment created the first noninvasive machinery to
successfully identify natural images based on brain patterns, and its
success spurred many more attempts to encode and decode neural
activities evoked by various cognitive tasks
[\citet{ReconSpeech}, \citet{GeneratingText}]. We focus only on the
prediction task, but note that gains in prediction would improve the
accuracy of identification as well. A complete description of the
experiment can be found in the supplementary materials of the original
paper [\citet{Kayetall}]. This is background for our work, which
begins in Section~\ref{secCorrInData}.

The data of this experiment are composed of the set of natural images,
and the fMRI scans recorded corresponding to the images. The images
were sampled from a library of gray-scale\vadjust{\goodbreak} photos depicting natural
scenes, objects, etc. Two \emph{nonoverlapping} random samples were
taken: the training sample (1750 images) was used for fitting the
models; and the validation sample ($m=120$ images) was used for
measuring the prediction accuracy. Images were sequentially displayed
in a randomized order, each image appearing multiple times ($n=13$).
BOLD contrasts---correlates of neural activity---were continuously
recorded by the fMRI along a~three-dimensional grid covering the visual
cortex, as the subject watched the images. For each voxel in the grid,
the responses were temporally discretized so that a single value (per
voxel) was associated with a single displayed image.
%
%
\begin{figure}

\includegraphics{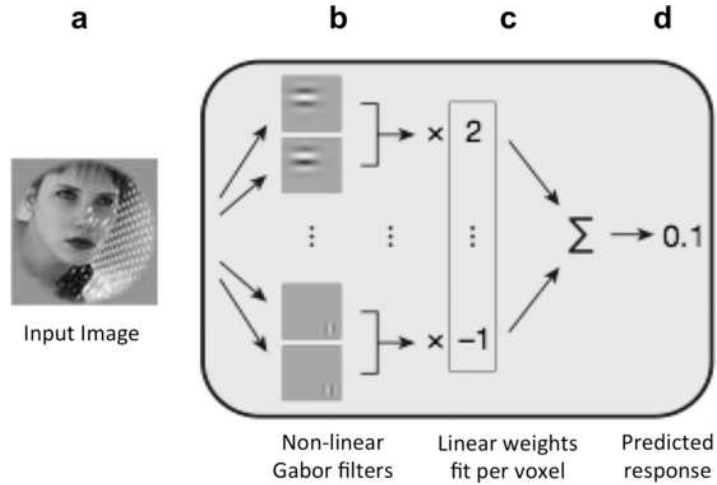}

\caption{A Gabor-based encoding model for natural
images. A cartoon depicting the encoding models that were used
for the fMRI experiment (adapted from Kay et al.). In these
models, natural images \textup{(a)} are
transformed into a vector of 10,409 features; features \textup{(b)}
measure the energy from two linear Gabor filters---tuned for specific
spatial frequency, location in the image and orientation---with
complementary phases. These features are combined according to a linear
weight vector \textup{(c)} that was fit separately for each voxel. The
predicted response of a voxel for an image is the weighted sum of the
features representing the image \textup{(d)}. The linear weights were
fit on the training data consisting of responses to 1750 images.}\label{Experiment}
\end{figure}
%
%
%
\begin{figure}

\includegraphics{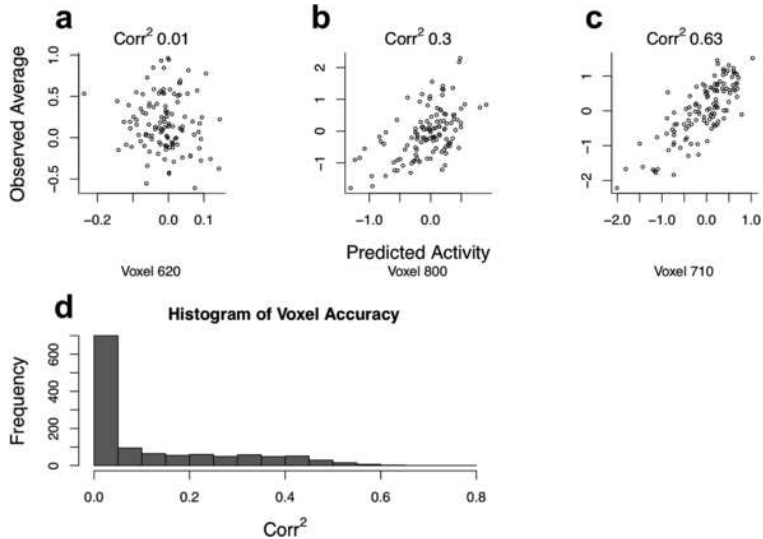}

\caption{Prediction accuracy for V1 voxels. Predicted
vs. observed average responses for three voxels in the V1 area,
reflecting poor \textup{(a)}, medium \textup{(b)} and high \textup{(c)} prediction accuracies.
Each point depicts the predicted response ($x$-axis) and the observed
average response (across all repeats) for an image of the validation
sample ($m=120$ images). \textup{(d)} Histogram of prediction accuracy for all 1250 V1 voxels.}\label{DataEA}
\end{figure}

Data from the training sample was used to fit a quantitative receptive
field model for each voxel, describing the fMRI response as a function
of the input image. Here is a brief overview; more details on these V1
encoding models can be found in \citet{VinceV1}. The model was
based on multiple Gabor filters capturing spatial location, orientation
and spatial-frequency of edges in the images (see
Figure~\ref{Experiment}). Because of the tuning properties of the Gabor
energy filters, this filter set is typically used for representing
receptive fields of mammalian V1 neurons. Gabor filters ($d={}$10,409
filters) transformed each image into a feature vector in $\mathbb
R^{d}$. For each of $Q$ voxels of interest, a linear weight vector
relating the features to the measurements was estimated from the
training data. Together, the transformation and linear weight vector
result in a \emph{prediction rule} that maps novel images to a~real-valued response per voxel.

In their paper, Kay et al. measured prediction accuracy by comparing
observations from the validation sample with the predicted responses
for those images. The validation data consisted of a total of $T=1560$
measurements (per voxel): $m=120$ different images, each repeated
$n=13$ times. Let $Y^{(r)}_t \in\mathbb{R}$ denote the measured fMRI activity
at voxel $r$ and time $t$. Repeated measurements of the same image were
averaged to reduce noise:
\[
\bar{Y}{}^{(r)}_j = \mathop{\operatorname{avg}}_{t\dvtx h(t)=j} Y^{(r)}_t, \qquad j=1,\ldots,m,
\]
where $h(t) \in\{1,\ldots, m\}$ indexes the image that was shown at
time $t$.

Let $f^{(r)}_{1},\ldots,f^{(r)}_{m}$ be the sequence of predictions for
voxel $r$, and $\bar{f}{}^{(r)}$ their average. A single value per
voxel summarizes prediction accuracy. That is,
\[
\operatorname{Corr}^2\bigl[\bigl(f^{(r)}_j
\bigr)_{j\leq m},\bigl(\bar{Y}{}^{(r)}_j
\bigr)_{j\leq m}\bigr]:= \frac{  (\sum_{j=1}^{m} (f^{(r)}_{j}-\bar{f}{}^{(r)})(\bar
{Y}{}^{(r)}_j-\bar{Y}{}^{(r)}) )^2} {
\sum_{j=1}^{m}(f^{(r)}_{j}-\bar{f}{}^{(r)})^2 \sum_{j=1}^{m}(\bar
{Y}{}^{(r)}_j-\bar{Y}{}^{(r)})^2}.
\]
In Figure~\ref{DataEA} we show examples of voxels with low,
intermediate and high prediction accuracies, and a histogram of the
accuracy for all $1250$ voxels located within the V1 area. We can drop
the superscript $r$ because each voxel is analyzed separately.
%

\subsection{Correlation in the data}\label{secCorrInData}
The goal of our work is to separate the two
factors that determine the accuracy of prediction rules: the adequacy
of the feature set or the linear model; and the noise level.
Explainable variance represents the optimal accuracy if prediction were
unrestricted by the choice of features and model.

We validate our explainable variance estimators by showing the
estimators account well for the differences in prediction accuracy
between $Q = 1250$ voxels within the primary visual cortex (V1). This
analysis depends on the assumption that the observed variation between
voxels is primarily due to differences in the level of the
signal-to-noise in the validation data, rather than, for example,
differences in the adequacy of the feature set underlying the
prediction models. Once the estimators are validated on this controlled
setting, explainable variance can be used more broadly, for example, to
compare the predictability levels of different functional areas.

\begin{figure}

\includegraphics{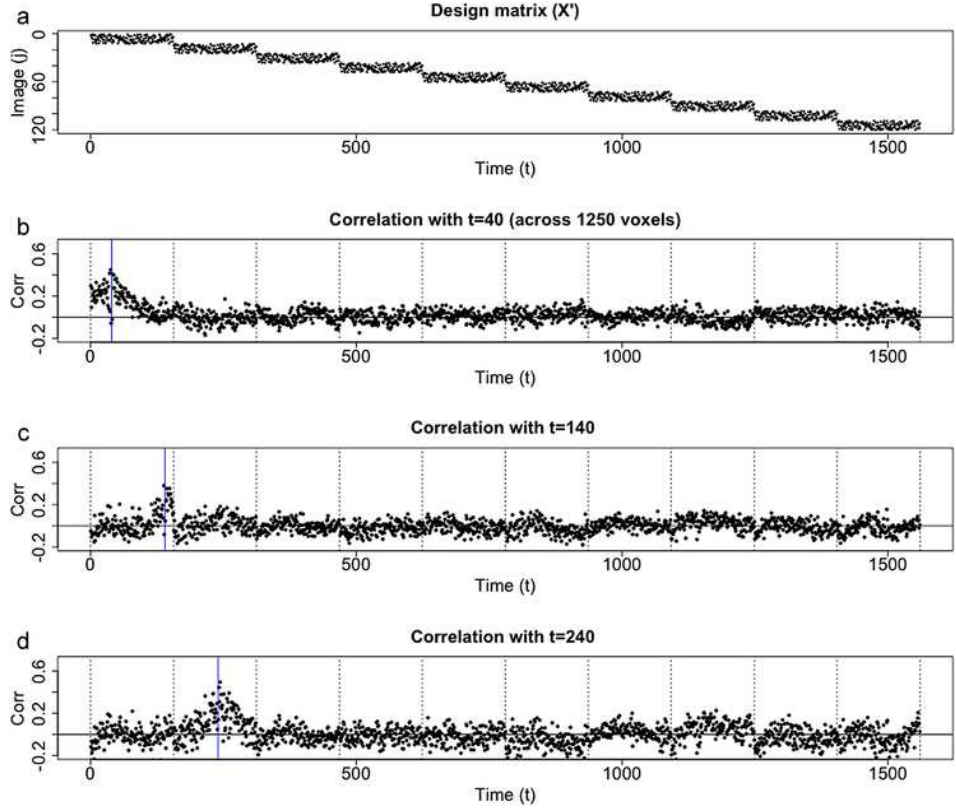}

\caption{Data acquisition for the validation data.
The responses in the validation set were collected in 10 separate
sessions (\emph{blocks}). Both
the design matrix for the experiment and the noise correlation are
influenced by this structure.
In \textup{(a)}, the transposed design matrix $X'$ is shown. This
matrix records which image ($y$-axis) is displayed at each time slot $t$
($x$-axis). Separate sets of 12 images were repeated $n=13$ times within
each block, whereas no image was recorded in more than one block. In
\textup{(b)--(d)} temporal autocorrelation is displayed, as measured
between a single time point [$t^* = 40, 140, 240$ for \textup{(b)},
\textup{(c)}, \textup{(d)}, resp.] and all others points. The $t = t^*$ point
is marked by a blue vertical line. On average, strong but nonsmooth
correlations are found within the blocks, but separate blocks seem
uncorrelated. Note that we calculate here the aggregate correlation of
the voxels, but cannot use this same method to measure the
autocorrelation of any specific voxel. Furthermore, the correlation
depicted here is due to both noise and signal.}\label{Covariance}
\end{figure}

Since we intend to use the validation data to estimate the explainable
variance, we now give a few more details on how it was collected.
Recall that the validation data consisted of $m = 120$ images each
repeated $n = 13$ times [see Figure~\ref{Covariance}(a)]. These data
were recorded in 10 separate sessions so that the subject could rest
between sessions; the fMRI was recalibrated at the beginning of each
session. Each session contained all presentations of 12 different
images. A pseudo-random integer sequence ordered the repeats within a
session.\footnote{The pseudo-random sequence allocated spots for 13
different images; no image was shown in the last category and the
responses were discarded.}

When we measure correlation across many voxels, it appears that the
design of the experiment induces strong correlation in the data. To see
this, in Figure~\mbox{\ref{Covariance}(b)--(d)} we plot the correlation between
measurements at different time slots (each time slot is represented by
the vector of $Q=1250$ measurements). This plot shows qualitatively the
correlation among individual voxels, including both noise driven and
possibly stimuli-driven correlations. Clearly, there are strong
correlations between time slots within a block, but no observable
correlations between blocks. As these within-block correlation patterns
do not correspond to the stimuli schedule, that is, randomized within a
block, we conclude the correlations are largely due to noise. These
noise correlations need to be taken into account to correctly estimate
the explainable variance.

%

\subsection{A probability model for the measurements}
\label{secSingleModel} We introduce a probabilistic model for the
measurements $\mathbf{Y}:= (Y_t)_{t=1}^T$ at a single voxel.
$\mathbf{Y}$ is modeled\vadjust{\goodbreak} as a random effects model with additive,
correlated noise [\citet{Williams52}]. We assume the observed
response is the sum of two independent random processes: the signal
process, which is random due to sampling of images into the validation
set, and the noise process describing fluctuations unrelated to the
stimuli. Additivity of noise is considered a good approximation for
fMRI-event related designs and is commonly used
[\citet{EfficientDesign}]. The random effects model accounts for
the generalization of prediction accuracy from the validation sample to
the larger population of natural images.

\subsubsection{Signal}
Images are shown in a long sequence, in which each of the $m$ images in
the random sample is repeated multiple times. The order of presentation
is described by the design matrix $X \in\{0,1\}^{T\times m}$ (see
Figure~\ref{Covariance}). Each row vector $X_t$ has a single 1 entry,
which identifies the image shown at time $t$,
\[
X_{t,j} = 1\qquad\mbox{if the $j$th image in the sample is displayed at time $t$}.
\]

For now, the sampling of images is conveyed through their effects on
the measurement. We use a homogenous set of random variables
[\citet{Pigeonhole}] to represent the mean responses to the images
in the sample. Let $A_j$ be the mean of the observed responses to the
$j$th image in the sample. We assume that
\[
A_j \sim\bigl(0,\sigma^2_A\bigr)\qquad\mbox{for } j=1,\ldots,m,
\]
that is, $A_j$'s are i.i.d. with mean $0$ and variance
$\sigma^2_A\geq0$, but are not necessarily normally distributed. In the
experiment we are analyzing, the randomness originates from sampling a
large (infinite) population of images. We use $\mathbf{A}:=
(A_j)_{j=1}^m$ for the random-effect column vector,\vspace*{-2pt}
$\bar{A}:=\frac{1}{m}\sum_{j=1}^m A_j$ for the sample mean, and $s^2_A:= \frac{1}{(m-1)}\sum_{j=1}^m (A_j-\bar{A})^2$ for the sample variance
of the random effects.

$X\mathbf{A}$ is the $T$-dimensional random signal vector, denoting the
random effect at each measurement. To index the effect corresponding to
time $t$, we use the shorthand $A_{(t)}:=X_t A$.

\subsubsection{Noise}
We assume the measurement noise process $\varepsilon
=(\varepsilon_t)_{t=1}^T$ is independent of the random signal vector
$X\mathbf{A}$. We further assume the noise elements $\varepsilon_t$
have 0 mean, $\sigma^2_\varepsilon\geq0$ variance, but may be
autocorrelated. The unknown correlation, denoted by a matrix
$\Sigma\in\mathbb{R}^{T \times T}
(\operatorname{diag}(\Sigma)=\mathbf{1})$, captures the slow-changing
hemodynamics of the BOLD and the effects of preprocessing on the BOLD
signals. Hence,
%
%
%
\begin{equation}
\label{DistEps} \mathbb{E}_\varepsilon[\varepsilon_t] = 0,\qquad
\operatorname{cov}(\varepsilon_t,\varepsilon _{u}) = \sigma^2_\varepsilon
\Sigma_{tu}, \Sigma_{tt}=1,
\end{equation}
or in matrix notation $\operatorname{cov}[\varepsilon] =
\sigma^2_\varepsilon\Sigma$.

\subsubsection{Model for observed responses}
We are now ready to introduce the observed response (column) vector
$\mathbf{Y} \in\mathbb R^T$ as follows:
%
%
%
\begin{equation}
\mathbf{Y} = X\mathbf{A}+\varepsilon
\end{equation}
and for a single time slot $t$
\[
Y_t = A_{(t)} + \varepsilon_t.
\]

\subsubsection{Response covariance}
The model involves two independent sources of randomness: the image
sampling, modeled by the random effects ($\mathbf{A}$), and the
measurement noise $\varepsilon$. Assuming independence between
$\mathbf{A}$ and $\varepsilon$, the covariance of $Y_t$ and $Y_u$
amounts to adding the individual covariances
%
%
%
\begin{eqnarray}
\operatorname{cov}_{A,\varepsilon}(Y_t, Y_u) &=& \operatorname{cov}_A(A_{(t)},
A_{(u)}) + \operatorname{cov}_\varepsilon (\varepsilon_t,
\varepsilon_u)
\nonumber\\[-8pt]\\[-8pt]
&=& \sigma^2_A 1_{(X_t=X_u)} +
\sigma ^2_\varepsilon \Sigma_{tu}.\nonumber
\end{eqnarray}
The first term on the RHS shows that treatment (random) effects are
uncorrelated if they are based on different inputs, but are identical
if based on the same input, with a variance of $\sigma_A^2$. In matrix
form, we get
%
%
%
\begin{equation}\label{covY}
\mathbb{E}_{A,\varepsilon}[\mathbf{Y}] = 0,\qquad \operatorname{cov}_{A,\varepsilon
}(\mathbf{Y})= \sigma^2_{A} XX'+ \sigma^2_\varepsilon
\Sigma.
\end{equation}

\subsection{Explainable variance and variance components}
\label{secQuantities} We are ready to define explainable variance, a
scaled version of the treatment variance $\sigma^2_A$. Explainable
variance is relevant to the performance of prediction models, a
property we will discuss in Section~\ref{secExplainable}.

Recall that $\bar{Y}_j$ are the averaged responses per image ($j =
1,\ldots,m$ for the images in our sample), and let $\bar{Y} = \frac{1}{T}
\sum_{t=1}^T Y_t$ be the global average response. Then the sample
variance of averages is
%
%
%
\begin{equation}
\mathrm{MS}_{\mathrm{bet}}:= \frac{1}{m-1}\sum_{j=1}^m
(\bar{Y}_j - \bar {Y} )^2.
\end{equation}
The notation $\mathrm{MS}_{\mathrm{bet}}$ refers to the mean-of-squares
between treatments. Let us define the total variance $\bar{\sigma}{}^2_Y$
as the population mean of $\mathrm{MS}_{\mathrm{bet}}$,
%
%
%
\begin{equation}
\bar{\sigma}{}^2_Y:= \mathbb{E}_{A,\varepsilon}[\mathrm{MS}_{\mathrm{bet}}].
\end{equation}
Note that $\bar{\sigma}{}^2_Y$ is not strictly the variance of any
particular $\bar{Y}_j$; indeed, the variance of $\bar{Y}_j$ is not
necessarily equal for different $j$'s.\footnote{In practice, this is
true for the individual measurements $Y_t$ as well. We chose
$\Sigma_{tt}=1$ for illustration reasons.} Nevertheless, we will
loosely use the term \emph{variance} here and later, owing to the
parallels between these quantities and the variances in the i.i.d.
noise case.

The average across repeats $\bar{Y}_j$ is composed of a signal part
($A_j$) and an average noise part ($\bar{\varepsilon}_j$); similarly,
$\bar{Y}$ is composed of $\bar{A}$ and $\bar{\varepsilon}$. By
partitioning the $\mathrm{MS}_{\mathrm{bet}}$, and taking expectations
over the sampling and the noise, we get
%
%
%
\begin{equation}
\qquad \mathbb{E}_{A,\varepsilon}[\mathrm{MS}_{\mathrm{bet}}] = \mathbb{E}_A\Biggl[
\frac{1}{m-1}\sum_{j=1}^m
(A_j-\bar{A} )^2\Biggr] +\mathbb{E}_\varepsilon
\Biggl[\frac{1}{m-1}\sum_{j=1}^m (
\bar{\varepsilon }_j - \bar {\varepsilon} )^2\Biggr],
\end{equation}
where the cross-terms cancel because of the independence of the noise
from the sampling. We can call the expectation of the second term the
\emph{noise level}, or $\bar{\sigma}{}^2_\varepsilon$, and get the
following decomposition:
%
%
%
\begin{equation}\label{vaY_decompose}
\bar{\sigma}{}^2_Y= \sigma^2_A +
\bar{\sigma}{}^2_\varepsilon.
\end{equation}
In other words, the signal variance $\sigma^2_A$ and the noise level
$\bar{\sigma}{}^2_\varepsilon$ are the signal and noise components of the
total variance.

Finally, we define the proportion of \emph{explainable variance} to be
the ratio
%
%
%
\begin{equation}\label{EVdef} \omega^2:= \sigma^2_A/\bar{\sigma}{}^2_Y.
\end{equation}
Explainable variance measures the proportion of variance due to signal
in the averaged responses; estimating it is the goal of this work. Note
that by definition, $\omega^2$~is restricted to $[0,1]$.

To estimate $\omega^2$, we need estimators for $\bar{\sigma}{}^2_Y$ and
$\sigma^2_A$. Whereas $\bar{\sigma}{}^2_Y$ can be directly estimated from
the sample, to estimate $\sigma^2_A$ we need a method to separate the
signal from the noise. In the following, we propose a method to
distinguish them using their different covariance structures. We first
develop some technical algebraic identities important for the
estimation procedure. Some readers might prefer to skip directly to
Section~\ref{secShuffle}.

\subsection{Quadratic contrasts}
\label{secQuadratic} In this subsection we express
$\mathrm{MS}_{\mathrm{bet}}$ as a quadra\-tic contrast of the full data
vector $\mathbf{Y}$. This contrast highlights the relation between
$\bar{\sigma}{}^2_Y$ or $\bar{\sigma}{}^2_\varepsilon$ with both the
design $XX'$ and the measurement correlations $\Sigma$, and produces
algebraic descriptions to be used for deriving the shuffle estimator.
These are simple extensions of classical treatment of variance
components [\citet{VarianceComponents}].

Denote $B:= XX'/n$, the $\mathbb{R}^{T\times T}$ matrix in which
%
%
%
\begin{equation}
B_{tu} = \cases{\displaystyle\frac{1}{n}, &\quad if $X_t =X_u$,
\vspace*{4pt}\cr
0, &\quad otherwise,}
\end{equation}
where $B$ is an averaging matrix, meaning that multiplication of a
measurement vector by $B$ replaces each element in the vector by the
treatment average, as in
%
%
%
\begin{equation}
(B\mathbf{Y})_t = \bar{Y}_{h(t)}.
\end{equation}
It is easy to check that $B = B'$ and $B = B^2$. Also, let
$G\in\mathbb{R}^{T\times T},   G_{tu} = 1/T $ for $ t,u = 1,\ldots,T$,
be the global average matrix, so that $(G\mathbf{Y})_t = \bar{Y},
t=1,\ldots,T$. We can now express $\mathrm{MS}_{\mathrm{bet}}$ as a
quadratic expression of $\mathbf{Y}$,
%
%
%
\begin{equation}
\mathrm{MS}_{\mathrm{bet}} = \frac{1}{(m-1)n} \bigl\|(B-G)\mathbf{Y}\bigr\|^2,
\end{equation}
or, more generally, as a quadratic function of any input vector,
%
%
%
\begin{equation}
\mathrm{MS}_{\mathrm{bet}}(\cdot):= \frac{1}{(m-1)n} \bigl\|(B-G) (\mathbf{\cdot})
\bigr\|^2.
\end{equation}

By replacing the $\mathrm{MS}_{\mathrm{bet}}$ with its quadratic form,
a relation is exposed between the total variance, the design and the
correlation of the noise:
\begin{eqnarray*}
\bar{\sigma}{}^2_Y&=& \mathbb{E}_{A,\varepsilon}
\bigl[\mathrm{MS}_{\mathrm{bet}}(\mathbf{Y})\bigr]
\\
&=& \frac{1}{(m-1)n} \mathbb{E}_{A,\varepsilon}\bigl[\operatorname{tr} \bigl( (B-G) \bigl(\mathbf{Y}'\mathbf{Y}
\bigr) (B-G) \bigr)\bigr]
\\
&=& \frac{1}{(m-1)n} \operatorname{tr} \bigl( (B-G) \operatorname{cov}_{A,\varepsilon}(\mathbf{Y}) \bigr).
\end{eqnarray*}
The signal effect and noise are additive, hence,
\[
\frac{1}{(m-1)n} \operatorname{tr} \bigl( (B-G) \operatorname{cov}_A(\mathbf{Y}) \bigr) +
\frac
{1}{(m-1)n} \operatorname{tr} \bigl( (B-G) \operatorname{cov}_{\varepsilon}(\mathbf{Y}) \bigr).
\]
Substituting $\operatorname{cov}_A(\mathbf{Y}) = n \sigma^2_{A} B$ and
$\operatorname{cov}_{\varepsilon}(\mathbf{Y}) = \sigma_{\varepsilon}^2
\Sigma$,
\begin{eqnarray*}
&& \frac{1}{(m-1)n} \operatorname{tr} \bigl( (B-G) \bigl(n
\sigma^2_A B \bigr) \bigr) + \frac {1}{(m-1)n} \operatorname{tr} \bigl(
(B-G) \sigma_{\varepsilon}^2 \Sigma \bigr)
\\
&&\qquad = \frac{1}{(m-1)}\sigma^2_A \operatorname{tr}(B-G) +
\frac{1}{(m-1) n} \sigma _{\varepsilon}^2 \operatorname{tr}\bigl((B-G)\Sigma\bigr)
\\
&&\qquad = \sigma^2_A + \frac{1}{(m-1) n} \sigma_\varepsilon^2\operatorname{tr} \bigl((B-G)\Sigma \bigr).
\end{eqnarray*}

%
\begin{derivation}\label{Prop1}
Under the model described in Section~\ref{secSingleModel},
%
%
%
\begin{equation}\label{MSB_Decompose}
\bar{\sigma}{}^2_Y= \sigma^2_A +
\frac{1}{(m-1) n} \sigma_\varepsilon ^2 \operatorname{tr} \bigl((B-G)\Sigma
\bigr).
\end{equation}
\end{derivation}

As a direct consequence of (\ref{vaY_decompose}) and (\ref{MSB_Decompose}),
we get an exact expression for the noise level
%
%
%
\begin{equation}
\label{vaY_express} \bar{\sigma}{}^2_\varepsilon=
\frac{1}{(m-1) n} \sigma_\varepsilon^2 \operatorname{tr} \bigl((B-G)\Sigma \bigr).
\end{equation}
This expression clarifies how $\bar{\sigma}{}^2_\varepsilon$ depends on
the design, the noise variance and the noise autocorrelation. As
expected, $\bar{\sigma}{}^2_\varepsilon$ scales linearly with the noise
variance of the individual measurements $\sigma^2_\varepsilon$. More
interesting is that $\bar{\sigma}{}^2_\varepsilon$ depends linearly on
$\operatorname{tr}(B \Sigma)$---the interplay between the design and
the noise autocorrelation.

Note that if the within treatment noise is uncorrelated, this
expression simplifies to a classical ANOVA result. Uncorrelated noise
within treatments manifests, in a~properly sorted version of $\Sigma$,
as small $n\times n$ identity blocks. Therefore, $\operatorname{tr}
((B-G)\Sigma ) = (m-1)\sigma^2_\varepsilon$ and
$\bar{\sigma}{}^2_\varepsilon= \sigma^2_\varepsilon /n$. In that case
$\bar{\sigma}{}^2_Y= \sigma^2_A + \sigma^2_\varepsilon/n$, and by
plugging in an estimator of $\sigma^2_\varepsilon$, we can directly
estimate $\bar {\sigma}{}^2_\varepsilon$ and $\sigma^2_A$. The estimator
for $\omega^2$ is
\[
\hat{\omega}{}^2 = 1 - \frac{1}{F}
\]
with $F$ being the standard $F$ statistic. This is the method-of-moments
estimator described fully in Section~\ref{secDataMethods}.

On the other hand, when some correlations within repeats are greater
than 0, $\sigma^2_\varepsilon/n$ underestimates the noise level and
inflates the explainable variance. In the next section we introduce the
shuffle estimators which can deal with correlated noise.

\section{Shuffle estimators for signal and noise variances}
\label{secShuffle} In this section we propose new estimators called the
shuffle estimators for the signal and noise levels, and for the
explainable variance. As in (\ref{vaY_decompose}), $\bar{\sigma}{}^2_Y=
\sigma^2_A + \bar{\sigma}{}^2_\varepsilon$, but the noise level
$\bar{\sigma}{}^2_\varepsilon$ is a function of the (unknown)
measurement correlation matrix $\Sigma$. Using shuffle estimators, we
can estimate $\sigma^2_A$ and $\bar{\sigma}{}^2_\varepsilon$ without
having to estimate the full $\Sigma$ or imposing unrealistically strong
conditions on it.

\begin{figure}

\includegraphics{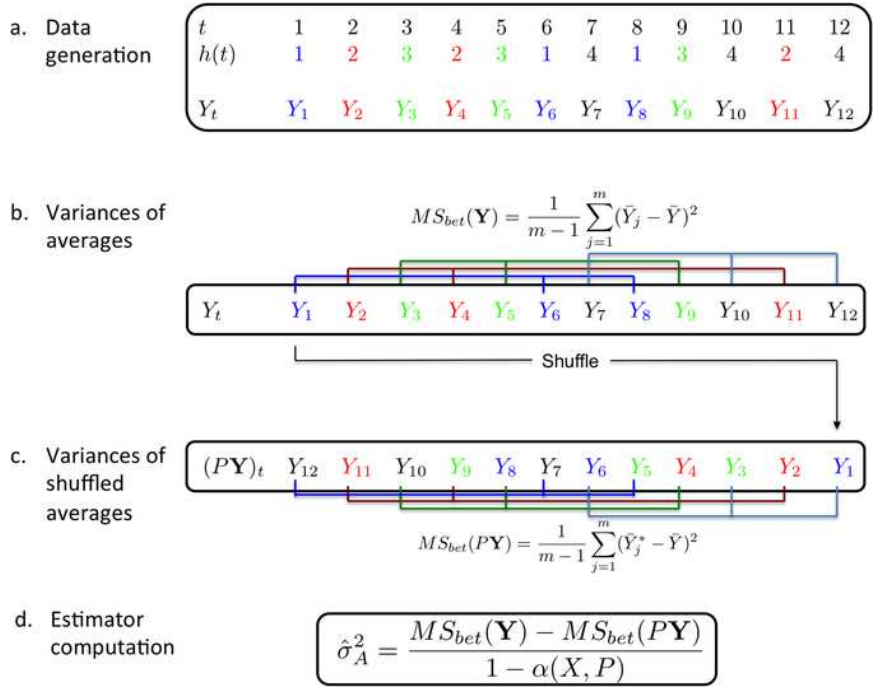}

\caption{Cartoon of the shuffle estimator.
\textup{(a)}~Data is generated according to a predetermined design, with each color
representing repeats of a different image.
\textup{(b)}~Repeats of each image are averaged together and the sample variance is computed on these averages.
\textup{(c)} Data is shuffled by $P$, in this example reversing the
order. Now measurements which do not originate from the same repeat are averaged
together ($\bar{Y}{}^*_j$'s), and the sample variance of the new averages
is computed. These averages should have a lower variance in expectation,
and we can calculate the reduction amount $\alpha= \frac{1}{m-1}
\operatorname{tr} ((B-G)PBP' )$, where $B=XX'/n$.
\textup{(d)}~The shuffle estimator for signal variance is the
difference between the two sample variances, after correction of $1-\alpha$.}\label{ShuffleCartoon}
\end{figure}

The key idea is to artificially create a second data vector that will
have similar noise patterns as our original data (see
Figure~\ref{ShuffleCartoon}). We do this by permuting, or
\emph{shuffling}, the original data along symmetries that we identify
in the data collection. In Section~\ref{secPerm} we formalize the
definition of such permutations that conserve the noise correlation,
and give plausible examples for neuroscience measurements. In
Section~\ref{secProof} we compare the variance of averages
($\mathrm{MS}_{\mathrm{bet}}$) of the original data
[Figure~\ref{ShuffleCartoon}(b)], with the same contrast computed on
the shuffled data (c). Because repeated measures for the same image are
shuffled into different categories, the variance due to signal will be
reduced in the shuffled data. We derive an unbiased estimator for
signal variance $\sigma^2_A$ based on this reduction in variance, and
use the plug-in estimators to estimate $\bar{\sigma}{}^2_\varepsilon$
and $\omega^2$.
%
%

\subsection{Noise-conserving permutation for $\mathbf{Y}$ with respect to design $X$} \label{secPerm}
A~prerequisite for the shuffle estimator
is to find a permutation that will conserve the noise contribution to
$\bar{\sigma}{}^2_Y$. We will call such permutations noise-conserving
w.r.t. to $X$.

Recall (\ref{vaY_express}),
\[
\bar{\sigma}{}^2_\varepsilon= \frac{1}{(m-1)n} \operatorname{tr}\bigl( (B-G )
\bigl(\sigma^2_\varepsilon \Sigma\bigr)\bigr),
\]
where $\sigma^2_\varepsilon\Sigma=
\operatorname{cov}_\varepsilon[\mathbf{Y}]$ as before. Let $P
\in\mathbb{R}^{T \times T}$ be a permutation matrix. Then we have the
following:

%
%
\begin{definition}
$P$ is noise conserving w.r.t. $X$, if
%
%
%
\begin{equation}
\operatorname{tr} \bigl( (B-G ) P \sigma^2_\varepsilon\Sigma P'
\bigr) = \operatorname{tr} \bigl( (B-G ) \sigma^2_\varepsilon\Sigma \bigr).
\end{equation}
Equivalently,
\[
\operatorname{tr} \bigl( (B-G ) \operatorname{cov}_\varepsilon[P \cdot\mathbf{Y}] \bigr) = \operatorname{tr} \bigl( (B-G )
\operatorname{cov}_\varepsilon[\mathbf{Y}] \bigr).
\]
\end{definition}

Although we define the noise-conserving property based on the
covariance, replacing the covariance with the correlation matrix
$\Sigma$ would not change the class of noise-conserving permutations.

We consider first two important cases of noise conservation. Though
most useful noise-conserving permutations will be derived from
assumptions regarding the noise correlation $\Sigma$, the
noise-conservation property can be derived from the design $X$ as well.
This can be seen in the first example.

\subsubsection{Trivial noise-conserving permutations}
A permutation $P$ that simply relabels the treatments is not a
desirable permutation, even though it is noise conserving. We call such
permutations trivial:

%
%
\begin{definition}
A permutation $P$, associated with permutation function $g_{P}\dvtx
\{1,\ldots,T\}\rightarrow\{1,\ldots,T\}$, is trivial if
%
%
%
\begin{equation}
X_t = X_u\quad\Rightarrow\quad X_{g_P(t)} =X_{g_P(u)}\qquad\forall t,u.
\end{equation}
\end{definition}

It is easy to show that a trivial $P$ does nothing, that is,
$\mathrm{MS}_{\mathrm{bet}}(P \mathbf{Y}) =
\mathrm{MS}_{\mathrm{bet}}(\mathbf{Y})$.

\subsubsection{\texorpdfstring{Noise-conserving permutations based on symmetries of $\Sigma$}
{Noise-conserving permutations based on symmetries of Sigma}}
\label{secCorrModels} A useful class of nontrivial noise-conserving
permutations is the class of symmetries in the correlation matrix
$\Sigma$: a symmetry of $\Sigma$ is a permutation $P$ such that
$P\Sigma P'=\Sigma$. If $P$ is a symmetry of $\Sigma$, then $P$ is
noise conserving regardless of the design. Here are three important
general classes of symmetries which are commonly applicable in
neuroscience:
\begin{enumerate}[3.]
\item[1.] \textit{Uncorrelated noise}. The obvious example is the
uncorrelated noise case $\Sigma=I$ where all responses are
exchangeable. Hence, any permutation is noise conserving.

\item[2.] \textit{Stationary time series}. Neuroscience data typically
    are recorded in a long sequence containing a large number of
    serial recordings at constant rates. It is natural to assume
    that correlations between measurements will depend on the time
    passed between the two measures, rather than on the location of
    the pair within the sequence. We call this the \emph{stationary
    time series}. Under this model $\Sigma$ is a Toeplitz matrix
    parameterized by $\{\rho_d\}_{d=0}^{T-1}$, the set of
    correlation values $\Sigma_{t,u} = \rho_d$, where $d = |t-u|$.
    Though the correlation values $\rho_d$'s are related, this
    parameterization does not enforce any structure on them. This
    robustness is important in the fMRI data we analyze. For this
    model, a permutation that \emph{reverses} the measurement
    vector is noise conserving
\[
(P\mathbf{Y})_t=\mathbf{Y}_{T+1-t}.
\]
This is the permutation we use on our data in
Section~\ref{secData}.

The shift operators of the form
$(P\mathbf{Y})_t=(\mathbf{Y})_{t+k}$ define a transformation that,
up to edge effects, can be considered noise conserving.

\item[3.] \textit{Block effect models}. Another important case is when
measurements are collected in distinct sessions or blocks.
Measurements from different blocks are assumed independent, but
measurements within the same block may be correlated, perhaps
because of calibration of the measurement equipment. We index
the block assignment of time $t$ with $\beta(t)$. A simple
parameterization for noise correlation would be to let
$\Sigma_{t,u} = \zeta(\beta(t), \beta(u))$ depend only on the
block identity of measurements $t$ and $u$. We call this the
\emph{block} structure. Under the block structure, any
permutation $P$ (associated with function $g_P$) that maintains
the identity of blocks, meaning
%
%
%
\begin{equation}
\beta(t) = \beta(u)\quad\Rightarrow\quad\beta\bigl(g_P(t)\bigr) = \beta
\bigl(g_P(u)\bigr),
\end{equation}
would be noise conserving w.r.t. any $X$.
\end{enumerate}

The scientist is given much freedom in choosing the permutation $P$,
and should consider both the variance of the estimator and the
estimator's robustness against plausible noise-correlation structures.
Establishing criteria for choosing the permutation $P$ is the topic of
current research.

\subsection{Shuffle estimators}
\label{secProof} We can now state the main results. From the following
lemma we observe that every noise-conserving permutation establishes a
linear mean equation with two parameters: $\sigma^2_A$ and $\bar
{\sigma}{}^2_\varepsilon$. The coefficient of
$\bar{\sigma}{}^2_\varepsilon$ is 1, whereas the coefficient for
$\sigma^2_A$ is the constant
%
%
%
\begin{equation}
\label{alphaeq} \alpha= \alpha(X,P) = \frac{1}{m-1} \operatorname{tr} \bigl( (B-G)
\bigl(PBP'\bigr) \bigr),
\end{equation}
which depends only on the design $B=XX'/n$ and the permutation $P$---both known to the scientist. The size of $\alpha$ reflects how well P
mixes the treatments; the greater the mix, the smaller $\alpha$.

%
%
\begin{lemma}
\label{lemma1} If $P$ is a noise-conserving permutation for
$\mathbf{Y}$, then
\begin{longlist}[(a)]
\item[(a)] $\mathbb{E}_{A,\varepsilon}[
    \mathrm{MS}_{\mathrm{bet}}(P\mathbf{Y})] = \alpha \sigma^2_A +
    \bar{\sigma}{}^2_\varepsilon$,

\item[(b)] $\alpha\leq1$, and the inequality is strict iff $P$ is
    nontrivial.
\end{longlist}
\end{lemma}

The proof of (a) involves derivations as in Section~\ref{secQuadratic}
[e.g., equation (\ref{MSB_Decompose})], and the proof of (b) further
requires the Cauchy--Schwarz inequality. Both are found in the supplementary materials
[\citet{SupplementShuffle}].

The consequence of part (b) is that for any nontrivial $P$, we get a
second mean equation, which is linearly independent from the equation
based on the original data (because $\alpha< 1$). In other words, for
a nontrivial $P$, the equation set
%
%
%
\begin{equation}
\label{MeanEq} \cases{ \mathbb{E}_{A,\varepsilon}\bigl[ \mathrm{MS}_{\mathrm{bet}}(
\mathbf{Y})\bigr] = \sigma^2_A + \bar{
\sigma}{}^2_\varepsilon,
\vspace*{2pt}\cr
\mathbb{E}_{A,\varepsilon}
\bigl[ \mathrm{MS}_{\mathrm{bet}}(P\mathbf{Y})\bigr] = \alpha\sigma^2_A+
\bar{\sigma}{}^2_\varepsilon}
\end{equation}
has a unique solution. Solving the two equations above, we arrive at an
unbiased estimator for the signal variance.

%
%
\begin{definition}
The shuffle estimator for the signal variance is defined as
%
%
%
\begin{equation}\label{eqsigma2a}
\hat{\sigma}{}^2_A = \frac{\mathrm{MS}_{\mathrm{bet}}(\mathbf{Y}) - \mathrm{MS}_{\mathrm{bet}}(P\mathbf
{Y})}{1-\alpha}.
\end{equation}
\end{definition}

Finally, we can plug in $\hat{\sigma}{}^2_A$ and $\hat{\bar{\sigma}}{}^2_Y
= \mathrm{MS}_{\mathrm{bet}}(\mathbf{Y})$ to get the \emph{shuffle
estimator} for the explainable variance $\omega^2$:
\[
\hat{\omega}{}^2 = \frac{ \hat{\sigma}{}^2_A}{\mathrm{MS}_{\mathrm{bet}}(\mathbf
{Y})} = \frac{1}{1-\alpha}
\biggl(1 - \frac{\mathrm{MS}_{\mathrm{bet}}(P\mathbf
{Y})}{\mathrm{MS}_{\mathrm{bet}}(\mathbf{Y})} \biggr).
\]

\begin{rems*}
\begin{enumerate}[3.]
\item[1.] An unbiased estimator of the noise level $\bar{\sigma
    }{}^2_\varepsilon$ can be derived from equations
    (\ref{eqsigma2a})~and~(\ref{vaY_decompose}),
\[
\hat{\bar{\sigma}}{}^2_\varepsilon= \mathrm{MS}_{\mathrm{bet}} - \hat{
\sigma}{}^2_A.
\]

\item[2.] Because $\hat{\sigma}{}^2_A$ estimates a nonnegative quantity,
    it is preferable to restrict $\hat{\sigma}{}^2_A$ to nonnegative
    values by taking $\hat{\sigma}{}^2_{A+} = \max \{0,
    \hat{\sigma}{}^2_{A} \}$.

\item[3.] The estimator is consistent under proper decay of the
    dependance. This statement is conditional on the asymptotic
    setup: explainable variance typically changes as the number of
    measurements ($T$) increases. Nevertheless, the shuffle
    estimator is consistent for a sequence of data sets (indexed by
    $k=1,2,\ldots $) of growing sizes $[T(k) \to\infty]$ for which
    total variance and explainable variance converge if (a) the
    number of treatments $m(k)$ grows to $\infty$ and (b) the
    dependence decays. For $\mathbf{Y}$ distributed as a
    multivariate gaussian,\footnote{More general SLLN conditions
    for the weakly dependent random variables
    $\bar{Y}{}^2_1(k),\ldots,\bar{Y}{}^2_{m(k)}(k)$ can be found in
    \citet{SLLNdependent}.} a sufficient condition for (b) can
    be given in terms of eigenvalues of $\Sigma$:
\begin{longlist}
\item[(b*)] The largest $m-1$ eigenvalues $\lambda_{(1)}(k),\ldots, \lambda_{(m-1)}(k)$ of the noise correlation
    matrix $\Sigma(k)$ satisfy
\[
\frac{1}{n^2 (m-1)^2}\sum_{i = 1}^{m-1}
\lambda_{(i)}^2(k) \to0\qquad\mbox{as } k \to\infty.
\]
\end{longlist}
Conditions (a) and (b*) assure that
$\operatorname{var}(\mathrm{MS}_{\mathrm{bet}}(\mathbf{Y}_k)) \to0$
as $k \to\infty$. The proof relies mainly on the expression for the
variance of a quadratic contrasts, as found, for example, in
\citet{LinearSearle}. For the proof of these results refer to
the supplementary materials [\citet{SupplementShuffle}].
\end{enumerate}
\end{rems*}

\section{Evaluating prediction for correlated responses}
\label{secExplainable} Although there are many uses for estimating the
explainable variance, we focus on its role in assessing prediction
models. \citet{Assessing} show that explainable variance upper
bounds the accuracy of prediction on the sample when noise is i.i.d. We
generalize their results for arbitrary noise correlation and account
for generalization from sample to population.\footnote{While these
results may have been proved before, we have not found them discussed
in similar context.} As shown in Lemma \ref{lemma2}, the noise level
$\bar{\sigma}{}^2_\varepsilon$ is the optimal expected loss under mean
square prediction error (MSPE) loss, and the explainable variance
$\omega^2$ approximates the accuracy under squared-correlation
$\operatorname{Corr}^2$ utility.

A more explicit notation setup is needed for studying the relation
between predictions and signal responses. Let $f$ be a prediction
function that predicts a real-valued response to any possible image
$I$, out of a large population of $M$ images,
%
%
%
\begin{equation}
f\dvtx  \{I_{i}\}_{i=1}^M \to\mathbb{R}.
\end{equation}
We will assume $f$ does not depend on the sample we are evaluating,
meaning that it was fit on separate data. We usually think of $f$ as
using some aspects of the image to predict the response, although we do
not restrict it in any parametric way to the image.

Prediction accuracy is measured only on the $m$ images sampled for the
(nonoverlapping) validation set. Let $\mathbf{s}\dvtx \{1,\ldots,m\}\to
\{1,\ldots,M\}$ be the random sampling function, and
$I_{\mathbf{s}(j)}$ the random image chosen for the $j$th sample image.

Furthermore, let us introduce notation relating the random effects to
the image sampling. For this, assume each image is associated with a
mean activation quantity $\mu_i$, so that $\sum\mu_i = 0$ and $\sum \mu_i^2 = \sigma^2_A$. Then the random effects $A_j$ defined before can
now be described $A_j = \mu_{\mathbf{s}(j)}$.

To evaluate prediction accuracy, the predicted response
$f(I_{\mathbf{s}(j)})$ is compared with the averaged (observed)
response for that image $\bar{Y}_j$. We consider two common accuracy
measures: mean squared prediction error ($\mathrm{MSPE}[f]$) and the squared
correlation ($\operatorname{Corr}^2[f]$), defined
%
%
%
\begin{eqnarray}\label{Corr2Define}
\mathrm{MSPE}[f]&:=& \frac{1}{m-1} \sum_{j=1}^m\bigl( f(I_{\mathbf{s}(j)})-\bar{Y}_{j} \bigr)^{2},
\\
\operatorname{Corr}^2[f]&:=& \operatorname{Corr}_{j}^{2} \bigl(
f(I_{\mathbf{s}(j)}), \bar{Y}_{j}\bigr)
\nonumber\\[-9pt]\\[-9pt]
&=& \frac{(1/(m-1)\sum_{j=1}^{m} (f(I_{\mathbf
{s}(j)})-\bar{f}_\mathbf{s})(\bar{Y_j}-\bar{Y}) )^2}
{1/(m-1)\sum_{j=1}^{m}(f(I_{\mathbf{s}(j)})-\bar{f}_\mathbf{s})^2 \sum_{j=1}^{m}(\bar{Y_j}-\bar{Y})^2},\nonumber
\end{eqnarray}
where $\bar{f_\mathbf{s}}$ denotes the average of the predictions for the
sample.

We will state and discuss the results relating the explainable variance
to optimal prediction; the proof can be found in the supplementary materials [\citet{SupplementShuffle}].

%
%
\begin{lemma}
\label{lemma2} Let $f^*\dvtx \{I_i\}_{i=1}^M \to\mathbb{R}$ be the
prediction function that assigns for each stimulus $I_i$ its mean
effect $\mu_i$, or $f^*(I_i) = \mu_i$. Under the model described in
Section~\ref{secSingleModel},
\begin{longlist}[(a)]
\item[(a)] $f^* = \arg\min_f \mathbb{E}_{A,\varepsilon}$ $[\mathrm{MSPE}[f]]$;

\item[(b)] $\bar{\sigma}{}^2_\varepsilon= \mathbb{E}_{A,\varepsilon
    }$ $[\mathrm{MSPE}[f^*]]$ $({=}\min_f \mathbb{E}_{A,\varepsilon}$ $[
    \mathrm{MSPE}[f]]$ by \textup{(a)});
\item[(c)] $\omega^2
    \approx\mathbb{E}_{A,\varepsilon}[\operatorname{Corr}^2[f^*]] $
    with a bias term smaller than $\frac{1}{m-1}$.
\end{longlist}
\end{lemma}

Under our random effects model, the best prediction (in MSPE) is
obtained by the mean effects, or $f^*$. More important to us, the
accuracy measures associated with the optimal prediction $f^*$ can be
approximated by signal and noise levels:
$\bar{\sigma}{}^2_\varepsilon$~for $\mathrm{MSPE}[f^*]$ and $\omega^2$ for $\operatorname{Corr}^2[f^*]$.

The main consequence of this lemma is that the researcher does not need
a ``good'' prediction function to estimate the ``predictability'' of
the response. Prediction is upper-bounded by $\omega^2$, a quantity
which can be estimated without setting a specific function in mind.
Moreover, when a researcher does want to evaluate a particular
prediction function $f$, $\hat{\omega}{}^2$ can serve as a yardstick with
which $f$ can be compared. If $ \operatorname{Corr}^2[f]
\approx\hat{\omega}{}^2$, the prediction error is mostly because of
variability in the measurement. Then the best way to improve prediction
is to reduce the noise by preprocessing\vspace*{1pt} or by increasing the number of
repeats. On the other hand, if $\operatorname{Corr}^2[f]
\ll\hat{\omega}{}^2$, there is still room for improvement of the
prediction function~$f$.

\section{Simulation}
\label{secSimulation} 
\begin{figure}[t!]

\includegraphics{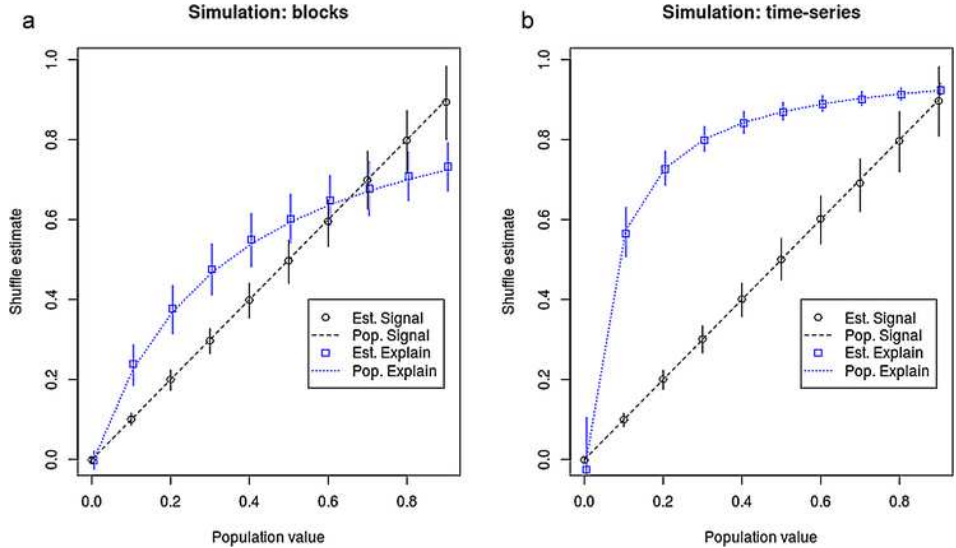}

\caption{Simulations for the block and time-series.
\textup{(a)}~Simulation results comparing shuffle estimates for signal
variance $\sigma^2_A$ (black) and explainable variance $\omega^2$
(blue) to the true population values (dashed line). Noise correlation
followed an independent block structure: noise within blocks was
correlated, and between blocks was independent. The $x$-axis represents
the true signal variance $\sigma^2_A$ of the data, and the $y$-axis marks
the average of the estimates and $[0.25,0.75]$ quantile range.
\textup{(b)}~A~similar plot for data generated under a stationary time-series model.}\label{Simulation}
\end{figure}

\begin{figure}[t!]

\includegraphics{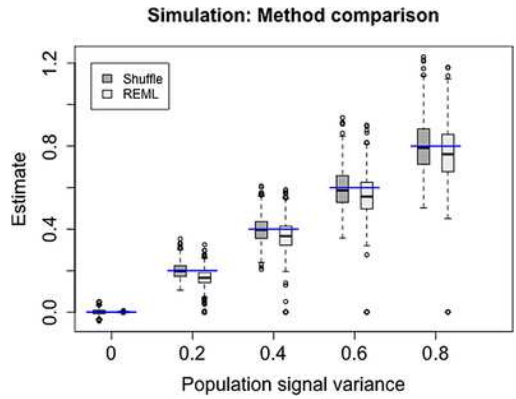}

\caption{Comparison of methods on simulation. Each pair
of box-plots represents the estimated signal variance $\sigma^2_A$
using the shuffle estimator (dark gray) and REML (light gray) for 1000
simulations. The blue horizontal line represents the true value of
$\sigma^2_A$. The REML estimator assumes the correct model for the
noise, while the shuffle estimator only assumes a stationary time
series. When there is no signal, REML outperforms the shuffle
estimators, but in all other cases it is both biased and has greater spread.}\label{SimComp}
\end{figure}

We simulate data with a noise component generated
from either a block structure or a times-series structure, and compute
shuffle estimates for signal variance and for explainable variance. For
a wide range of signal-to-noise regimes, our method produces unbiased
estimators of $\sigma^2_A$. These estimators are fairly accurate for
sample sizes resembling our image-fMRI data, and the bias in the
explainable variance $\omega^2$ is small compared to the inherent
variability. These results are shown in Figure~\ref{Simulation}. In
Figure~\ref{SimComp} we show that under nonzero $\sigma^2_A$, the
shuffle estimates have less bias and lower spread compared to the
parametric model using the correctly specified noise correlation.
%
%
%
%

\subsection{Block structure}
For the block structure we assumed the noise is composed of an additive
random block effect constant within blocks ($b_k$, $k=1,\ldots,B$
blocks), and an i.i.d. Gaussian term ($e_t$, $t=1,\ldots,T$)
\[
Y_t = A_{(t)} + b_{\beta(t)} + e_t.
\]
$A_j, b_k $ and $ e_t$ are sampled from centered normal distributions
with variances ($\sigma^2_A,\sigma^2_b,\sigma^2_e$). We used
$\sigma^2_b = 0.5, \sigma^2_e = 0.7$, and varied the signal level
$\sigma^2_A = 0, 0.1,\ldots,0.9$. We used $m=120$, $n=15$, with all
presentations of every 5 stimuli composing a block ($B=20$ blocks). For
each of these scenarios we ran 1000 simulations, sampling the signal,
block and error effects. $\mathrm{MS}_{\mathrm{bet}}$ was estimated the
usual way, and $P$ was chosen to be a random permutation within each
block ($\alpha= 0.115 $). The results are shown in
Figure~\ref{Simulation}(a).

\subsection{Time-series model}
\label{secTSMod} For the time-series model we assumed the noise vector
$e \in\mathbb{R}^T$ is distributed as a multivariate Gaussian with mean
0 and a covariance matrix $\Sigma$, where $\Sigma$ is an exponentially
decaying covariance with a nugget,
\[
\Sigma_{tu} = \rho_{|t-u|} = \lambda_1\cdot \exp
\bigl\{-|t-u|/\lambda_2\bigr\} + (1-\lambda_1)1_{(t=u)}.
\]
Then $Y = A_{(t)} + e_t$ with the random effects $A_{(t)}$ sampled from
$\mathcal{N}(0,\sigma^2_A)$ for $\sigma^2_A = 0, 0.1,\ldots, 0.9$. We
used $m=120, n=15$, and the parameters for the noise were $\lambda_1 =
0.7$ and $\lambda_2 = 30$, meaning $\rho_{125} \approx0.01$. The
schedule of treatments was generated randomly. For each of these
scenarios we ran 1000 simulations, sampling the signal and the noise.
In Figure~\ref{Simulation}(b) we estimated the shuffle estimator with $P$
the reverse permutation ($g_P(t) = T+1-t$), resulting in $\alpha=
0.064$.

\subsection{Comparison to REML}
In Figure~\ref{SimComp} we used time-series data to compare
$\sigma^2_A$ estimates based on the shuffle estimators to those
obtained by an REML estimator with the correct parametrization for the
noise correlation matrix. We used the \texttt{nlme} package in R to fit
a repeated measure analysis of variance for the exponentially decaying
correlation of noise with a nugget effect. The comparison included 1000
simulations for $\sigma^2_A = 0, 0.2, 0.4, 0.6, 0.8$ and a noise model
identical to Section~\ref{secTSMod}.

\subsection{Results}
Figure~\ref{Simulation} describes the performance of shuffle estimates
on two different scenarios: block correlated noise (a), and stationary
time-series noise (b). For signal variance (black) the shuffle
estimator gives unbiased estimates. The shuffle estimator for
explainable variance is not unbiased, but the bias is negligible
compared to the variability in the estimates. In Figure~\ref{SimComp}
we compare the signal variance estimates based on the shuffle estimator
(dark gray) with estimates based on REML (light gray). The estimates
based on the shuffle have no bias, while those based on REML
underestimate the signal. The variance of the REML estimates is
slightly larger, due in part to rare events (between 1\%--0.5\% of runs)
in which the estimated signal variance was effectively 0---perhaps
indicating a problem with the optimization.

\section{Data}
\label{secData} We are now ready to evaluate prediction models using
the shuffle estimates for explainable variance. Prediction accuracy was
measured for encoding models of 1250 voxels within the primary visual
cortex (V1). Because V1 is functionally homogenous, encoding models for
voxels within this cortical area should work similarly. As observed in
Figure~\ref{DataEA}, there is large variation between prediction
accuracies for the different voxels. We try to explain this observed
variation as a result of variation in the explainable variance. To do
this, prediction accuracy values for these 1250 voxels are compared to
the explainable variance estimates generated by the shuffle estimator
for each voxel. We also compare the accuracy values to alternative
estimates for explainable variance, using the method of moments for
uncorrelated noise, and REML under several parameterizations for the
noise.

\subsection{Methods}
\label{secDataMethods} We estimate the explainable variance of voxels
($\omega^2 =\break  \sigma^2_A/\bar{\sigma}{}^2_Y$) with several different
methods. The methods differ in how $\sigma^2_A$ is estimated; all
methods use the sample averages variance
$\mathrm{MS}_{\mathrm{bet}}(\mathbf{Y})$ for $\bar{\sigma}{}^2_Y$ and
plug in the two estimates into $\omega^2$. We estimate $\omega^2$
separately for each voxel ($r = 1,\ldots,1250$). The methods we compare
are as follows:
\begin{enumerate}[3.]
\item[1.] The shuffle estimator. We assume the noise follows a
stationary time-series model within each block and is
independent between the blocks. We therefore choose a
permutation $P$ that reverses the order of the measurements,
$(P\mathbf{Y})_t=\mathbf{Y}_{T+1-t}$. Because the size of the
blocks is identical, reversing the order of the data vector is
equivalent to reversing the order within each block, $\alpha=
0.14$. Specifically, the estimator is restricted to be
positive:
\[
\hat{\sigma}{}^2_{A+} = \max \biggl\{\frac{\mathrm{MS}_{\mathrm{bet}}(\mathbf{Y}) -
\mathrm{MS}_{\mathrm{bet}}(P\mathbf{Y})}{1- \alpha}, 0
\biggr\}
\]
for signal variance, and $\hat{\omega}{}^2 = \hat{\sigma}{}^2_{A+} /
\mathrm{MS}_{\mathrm{bet}}$ for the explainable variance.

\item[2.] An estimator ($\tilde{\omega}{}^2$) unadjusted for correlation.
    We use the mean square within ($\mathrm{MS}_{\mathrm{wit}} =
    \frac{1}{(m-1)n} \sum_{j=1}^m \sum_{t\dvtx h(t)=j} (Y_t -
    \bar{Y}_j)^2$) contrast to estimate the noise variance
    $\sigma^2_\varepsilon$, scale by $1/n$ to estimate the noise
    level $\bar{\sigma}{}^2_\varepsilon$, and remove the scaled
    estimate from $\mathrm{MS}_{\mathrm{bet}}$,
\[
\tilde{\sigma}{}^2_A = \mathrm{MS}_{\mathrm{bet}} -
\mathrm{MS}_{\mathrm{wit}}/n.
\]
Explainable variance is obtained by plug-in estimator
$\tilde{\omega}{}^2 = \tilde{\sigma}{}^2_A/\mathrm{MS}_{\mathrm{bet}}$.

\item[3.] Estimators based on a parametric noise model.
\begin{itemize}
\item We assume the noise is generated from an exponentially
decaying correlation matrix, with a nugget effect. This
means
\[
\Sigma_{t,t+d} = \lambda_{2}\exp(-d/\lambda_1)+1_{(d=0)}(1-
\lambda_2),
\]
where the rate of decay $\lambda_1$ and nugget effect
$\lambda_2$ were additional parameters. If $\lambda_2 =0$,
this is equivalent to the $\operatorname{AR}(1)$ model.

\item Alternatively, we assume the noise is generated from an
    $\operatorname{AR}(3)$ process or $\varepsilon_t = \eta_t +
    \sum_{k=1}^3a_k\varepsilon_{t-k}$. This models allows for
    nonmonotone correlations.
\end{itemize}
We use the \texttt{nlme} package in R to estimate the signal
variance of this model using restricted maximum likelihood [REML,
e.g., \citet{REML}], and use the plug-in estimator for the
explainable variance.
\end{enumerate}

%
%
\begin{figure}[b!]

\includegraphics{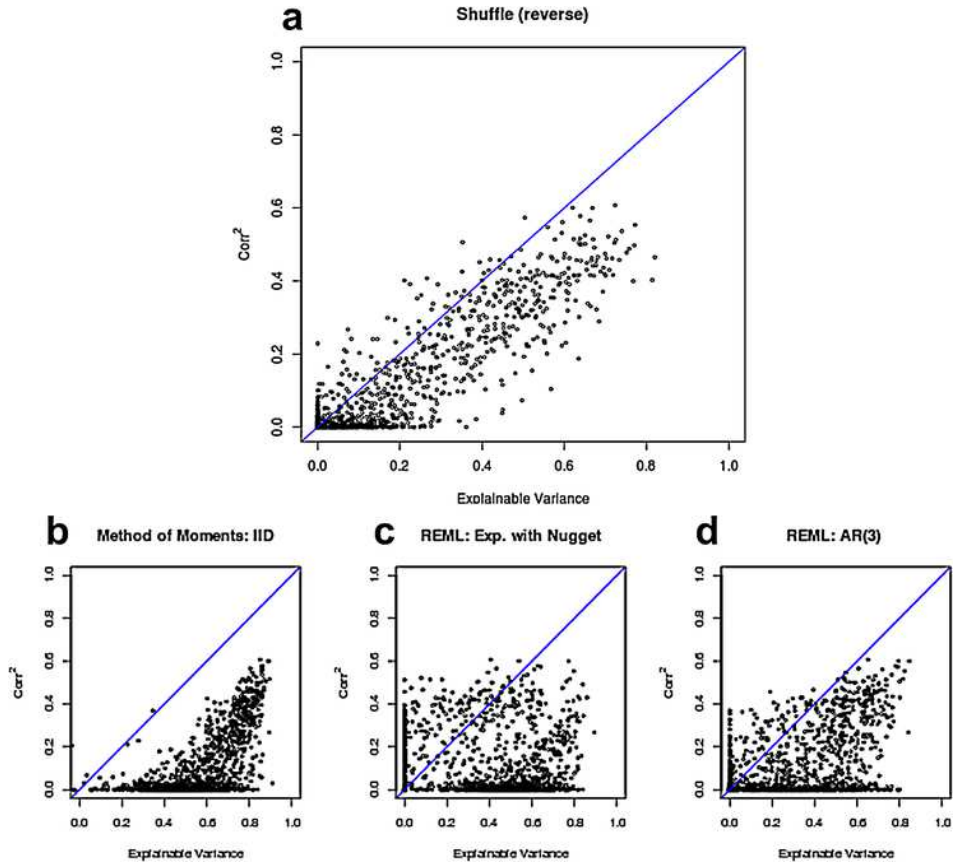}

\caption{Optimal vs. observed prediction accuracy. The
estimated optimal prediction is compared with observed prediction
($\operatorname{Corr}^2$), each point representing a V1 response. The optimal
prediction estimated by \textup{(a)} shuffle estimators accounting for
stationary noise distributions; \textup{(b)} Method of moments
estimator assuming independent noise; \textup{(c)} REML estimator
assuming exponential decay of noise with nugget within blocks; and
\textup{(d)} REML estimator assuming an $\operatorname{AR}(3)$ model for the noise
correlation within blocks. The $x=y$ is plotted in blue.}\label{OptvsObs}
\end{figure}

%
\subsection{Results}
In Figure~\ref{OptvsObs} we compare the prediction accuracy of the
voxels to estimates of the explainable variance. Each panel has 1250
points representing the 1250 voxels: the $x$ coordinate is the estimate
of explainable variance for the voxel, and the $y$ coordinate is
$\operatorname{Corr}^2[f]$ for the Gabor-based prediction rule. The
large panel shows the shuffle estimators for explainable variance. The
relation between $y = \operatorname{Corr}^2[f]$ and $\hat{\omega}{}^2$ is
very linear ($r=0.9$). The estimated slope and intercepts (using least
squares) for this relation are $y \approx0.66 \cdot\hat{\omega}^2 -
0.009$. Note that almost all voxels for which accuracy is close to
random guessing ($\operatorname{Corr}^2[f]<0.05$) could be identified
based on low explainable variance without knowledge of the specific
feature set.

When we try to repeat this analysis with other $\omega^2$ estimators,
explainable variance estimates are no longer strongly related with the
prediction accuracy. When correlation in the noise is ignored (b),
signal strength is greatly overestimated. In particular, some of the
voxels for which prediction accuracy is almost 0 have very high
estimates of explainable variance (as high as $\tilde{\omega}{}^2 =
0.8$). In contrast to the shuffle estimates, it is hard to learn from
these explainable variance estimates about the prediction accuracy for
a voxel.

This incompatibility of prediction accuracy and explainable variance
estimates is also observed when the estimates are based on maximum
likelihood methods that parameterize the noise matrix. For the
$\operatorname{AR}(3)$ model in (d), we see variability between explainable variance estimates
for voxels with given prediction accuracy level. The smaller model (c)
seems to suffer from both overestimation of signal and from high
variance.

\section{Discussion}\label{secDiscussion}
We have presented the shuffle estimator, a
resampling-based estimator for the explainable variance in a
random-effects additive model with autocorrelated noise. Rather than
parameterize and estimate the correlation matrix of the noise, the
shuffle estimator treats the contribution of the noise to the total
variance as a single parameter. Symmetries in the data collection
process indicate those permutations which, when applied to the original
data, would not change the contribution of the noise. An unbiased
estimator of the signal variance is derived from differences between
the total variance of the original data vector and the shuffled vector.
The resulting estimate of signal variance is plugged in as the
enumerator for the explainable variance ratio estimate.

For a brain-encoding experiment, we have shown that the strong
correlation present in the fMRI measurements greatly compromises
classical methods for estimating explainable variance. We used
prediction accuracy measures of a well-established parametric model for
voxels in the primary visual cortex as correlates of signal variance at
each of the voxels. Shuffle estimates explained most of the variation
in prediction accuracy between voxels, even though they were blind to
features of the image. Other methods did not do well: methods that
ignored noise correlation seem to greatly overestimate the explainable
variance, while methods that estimated the full correlation matrix were
considerably less informative with regards to prediction accuracy. We
consider this convincing evidence that the shuffle estimators for
explainable variance can be used reliably even when no gold-standard
prediction model is present.

Explainable variance is an assumption-less measure of signal, in that
it makes no assumptions about the structure of the mean function that
relates the input image to response. We find it attractive that the
shuffle estimator for explainable variance similarly requires only weak
assumptions for the correlation of the noise. This makes the shuffle
estimator a robust tool, which can be used at different stages of the
processing of an experiment: from optimizing of the experimental
protocol, through choosing the feature space for the prediction models,
to fitting the prediction models.

\subsection*{Choice of permutations}
The bias and variance of the shuffle estimator depend on the
permutation underlying the shuffle. Different permutations pose
different assumptions on the noise correlation as well as provide a
different mix of treatments corresponding to different $\alpha$s. We
recommend the permutation be chosen prior to the analysis, based on the
expected noise structure and the mixing constants ($\alpha$s), to
minimize the risk of data snooping. Optimally, the experiment could be
designed so that a specific permutation---perhaps the reverse
permutation---will mix treatments well, resulting in a low $\alpha$.
In our experiment, the reverse and other regular permutations such as
shifts had low $\alpha$s because the design was generated using an
irregular pseudo-random sequence. Moreover, when several
noise-conserving permutations exist and have similar $\alpha$s, it may
be preferable to average the corresponding shuffle estimators to reduce
the variance of the estimators.

In cases were no symmetry permutations are useable, a wider class of
``almost noise-conserving'' permutations can be considered. To give
concrete examples, consider the following two permutations: A cyclic
left-shift permutation so that $(P_1 \mathbf{Y})_t = \mathbf{Y}_{t+1}$
for $t = 1,\ldots,T-1$, and $(P_1 \mathbf{Y})_T =
\mathbf{Y}_1$; and a permutation of odd and even channels
$(P_2\mathbf{Y})_{2s-1} = \mathbf{Y}_{2s}$, $(P_2\mathbf{Y})_{2s} =
\mathbf{Y}_{2s-1}$, for $s = 1,\ldots,T/2$. Neither is an exact
symmetry of a covariance matrix that represents a stationary time
series. In $P_1$, the first measurement in each block is not correlated
to the sequence. $P_2$ is even farther from symmetry, in that the
medium and long-range correlations are conserved but the local
structure is scrambled.

Nevertheless, the shuffle estimates from either $P_1$ or $P_2$ produce,
when compared to predictions, population results that are similar to
those observed for the reverse permutation. The new estimates compare
in both linearity, with $r = 0.905$ between
$\operatorname{Corr}^2[f^*]$ and $\hat{\omega}{}^2$ for both $P_1$ and
$P_2$ compared to $r=0.9$ for the reverse permutation, as well in the
slope (0.65 for $P_1$, 0.68 for $P_2$) of the linear trend and its
intercept (0 for both). This does not imply that any permutation would
work well; indeed, shuffling with a random permutation, implicitly
assuming i.i.d. noise, results in a similar bias as that observed in the
method of moments estimator shown in Figure~\ref{OptvsObs}(b). Note
that for any candidate $P$ and possible $\Sigma$, the degree to which
noise is conserved can be explicitly measured by comparing
$\operatorname{tr}(\Sigma(B-G))$ to $\operatorname{tr}(P'\Sigma
P(B-G))$.

The shuffle estimators may be useful for applications outside of
neuroscience. These estimators can be used to estimate the variance
associated with the treatments of an experiment, conditioned on the
design, whenever measurement noise is correlated. Spatial correlation
in measurements arise in many different domains, from agricultural
experiments to DNA microarray chips. Shuffle estimators could provide
an alternative to parametric fitting of the noise contributions for
these applications.

Future research should be directed at expressing the variance of the
shuffle estimator for a candidate permutation, as well as at developing
optimal ways to combine information from multiple noise-conserving
permutations. More generally, shuffle estimators are a single example
of adapting relatively new nonparametric approaches from hypothesis
testing into estimation; we see much room for expanding the use of
permutation methods for creating robust estimators for experimental
settings.

\section*{Acknowledgments}
Y. Benjamini gratefully acknowledges support from the NSF VIGRE
fellowship. We are grateful to An Vu and members of Jack Gallant's
laboratory for access to the data and models and for helpful
discussions about the method, and to Terry Speed and Philip Stark for
suggestions that greatly contributed to this work. We are also thankful
to two anonymous reviewers and two Editors whose insightful comments
helped improve this manuscript.

\begin{supplement}
\stitle{Supplementary material}
\slink[doi]{10.1214/13-AOAS681SUPP} 
\sdatatype{.pdf}
\sfilename{aoas681\_supp.pdf}
\sdescription{We provide proofs for
Lemmas \ref{lemma1} and \ref{lemma2} in the supplementary, as well as describe and prove conditions for
consistency of the shuffle estimator.}
\end{supplement}


%

\printaddresses

\end{document}